\documentclass{jfm}
\usepackage{graphicx}
\usepackage{epstopdf, epsfig}
\usepackage{amsmath}
\usepackage{hyperref}
\hypersetup{colorlinks,
citecolor=blue}

\shorttitle{Non-interacting gravity waves}
\shortauthor{N. S. Ussembayev}

\title{Non-interacting gravity waves on the surface of a deep fluid\aff{\thanks{\bf{Dedicated to Professor Vladimir Zakharov's 80th Birthday}}}}

\author{Nail S. Ussembayev
  \corresp{\email{nussemba@indiana.edu}}}

\affiliation{\aff{}Computer, Electrical, and Mathematical Sciences and Engineering Divison\\
 King Abdullah University of Science and Technology, Thuwal 23955-6900, KSA}

\begin{document}
\maketitle

\begin{abstract}
We study the interaction of gravity waves on the surface of an infinitely deep ideal fluid. Starting from Zakharov's variational formulation for water waves we derive an expansion of the Hamiltonian to an arbitrary order, in a manner that avoids a laborious series reversion associated with expressing the velocity potential in terms of its value at the free surface. The expansion kernels are shown to satisfy a recursion relation enabling us to draw some conclusions about higher-order wave-wave interaction amplitudes, without referring to the explicit forms of the individual lower-order kernels. In particular, we show that unidirectional waves propagating in a two-dimensional flow do not interact nonlinearly provided they fulfill the energy-momentum conservation law. Switching from the physical variables to the so-called normal variables we explain the vanishing of the amplitudes of fourth- and certain fifth-order non-generic resonant interactions reported earlier and outline a procedure for finding the one-dimensional wave vector configurations for which the higher order interaction amplitudes become zero on the resonant hypersurfaces. 
\end{abstract}


\vspace{-1cm}

\section{\label{sec:intro} Introduction}
 
Many completely integrable nonlinear equations possessing beautiful mathematical structure arise in the study of water waves at various levels of approximation. 
The Korteweg-de Vries (KdV) and Camassa-Holm equations modeling long waves propagating in shallow water, the Sch\"{o}dinger equation with a cubic nonlinearity describing the evolution of a one-dimensional envelope of a wave train in a fluid of finite or infinite depth are just a few well-known members of a relatively large family of exactly solvable initial-value problems that can be derived from the set of equations underlying an ideal fluid with a free surface subject to a vertical gravitational force \citep{J}. 

In 1994  \cite{DZ1} conjectured that not only a few interesting limiting cases but also the fully nonlinear Euler equations for the potential flow of an ideal fluid with a free surface might be completely integrable under certain assumptions. Specifically, they considered weakly nonlinear waves on the surface of infinitely deep water and computed amplitudes of four-wave resonant interactions within a pure gravity wave spectrum in one horizontal dimension. According to \citet{P} quartic resonances are the first to occur for deep-water gravity waves as three-wave processes are forbidden by the corresponding dispersion relation. Surprisingly, the fourth order non-generic wave-wave interaction amplitude vanished identically on the resonant hypersurface. This unexpected finding was a strong indication that the resulting approximate system is integrable and suggested that the original hypothesis of integrability of the free-surface hydrodynamics equations - or its weaker counterpart asserting the existence of an additional integral of motion (besides mass, energy, and momentum) - is plausible \citep{ZS1, ZS2}. 

Employing the technique of Birkhoff normal form transformation \citet{CW}  independently verified conclusions of \cite{DZ1} and carried out the computations to the next order in nonlinearity discovering the resonant fifth order interactions with nonzero amplitudes. Subsequently, \citet{DLZ} found a compact expression for a certain fifth order amplitude and more recently showed that the interaction coefficient is nonzero for a family of resonant sextets \citep{DKZ}. These calculations are sufficient to deduce that the associated truncated systems are non-integrable. 

The fact that the cancellation occurring at the fourth order of perturbation is not accidental was confirmed by \citet{L} who derived all the possible topologically different wave vector configurations for the resonant five-wave interaction in one horizontal dimension and showed that for some orientations of the wave vectors the interaction coefficients are zero and for others, they are remarkably simple and compact. Nonetheless, the precise reason for the vanishing of the kernels for peculiar alignments of the wave vectors remained unexplained till date.   

As the order of nonlinearity increases the calculations become rather lengthy requiring at times summation of thousands of terms. This presents a serious obstacle even with today's computer capacities. In this article, we present an efficient way of calculating the expansion kernels that can serve as an avenue for exploring new integrable systems or discovering new integrals of motion and performing analytical computations and numerical simulations that used to be infeasible due to the enormous complexity of the involved calculations.  


By focusing on the unidirectional gravity waves in a two-dimensional flow we show that waves initially propagating in the same direction do not interact resonantly to generate waves propagating in the opposite direction. Interestingly, this result is not specific to the surface waves on deep water: as demonstrated recently by \citet{HN}, the nonlinear Alfv\'{e}n waves in an incompressible MHD plasma share the same property, and perhaps this fact too can be traced to the Hamiltonian structure \citep{ZK} of the governing set of equations.

\section{\label{sec:sec2} Hamiltonian theory of weakly nonlinear surface waves}

\subsection{\label{sec:sec2a} Hamiltonian and its expansion to arbitrary order}

The potential flow of an incompressible inviscid fluid characterized
by the velocity field ${\bf v}=\left(\nabla\phi,\phi_{z}\right)$,
vanishing with increasing depth, obeys the Laplace's equation 
\begin{equation}\label{lap}
\Delta\phi+\phi_{zz}=0
\end{equation}
in the bulk of the fluid together with the kinematic and dynamic boundary
conditions at the free surface $\xi({\bf x},t)$:

\begin{subequations}\label{eq1}
\begin{align}
\xi_{t}+\nabla\phi\cdot\nabla\xi-\phi_{z}=0,\label{eq1a}\\
\phi_{t}+\frac{1}{2}\left|\nabla\phi\right|^{2}+\frac{1}{2}\phi_{z}^{2}+g\xi=0\label{eq1b}
\end{align}
\end{subequations}
where ${\bf x}$ is the lateral coordinate, $\nabla$ is the corresponding
gradient operator and $g$ is the acceleration due to gravity. The
two conditions \eqref{eq1a}-\eqref{eq1b} at the surface are equivalent to the Hamilton's
equations of motion with canonically conjugate variables $\xi({\bf x},t)$
and $\psi({\bf x},t)\equiv\phi\left({\bf x},z=\xi,t\right)$: 
\[
\frac{\partial\xi}{\partial t}=\frac{\delta H}{\delta\psi},\ \ \ \frac{\partial\psi}{\partial t}=-\frac{\delta H}{\delta\xi}
\]
where $\delta$ denotes the variational derivative and $H\left(\xi,\psi\right)$
is the Hamiltonian functional 
\[
H=\frac{1}{2}\int_{\mathbb{R}^{d}}d{\bf x}\int_{-\infty}^{\xi}\left(\left|\nabla\phi\right|^{2}+\phi_{z}^{2}\right)dz+\frac{g}{2}\int_{\mathbb{R}^{d}}\xi^{2}d{\bf x}
\]
representing the total energy of the wave motion \citep{Z}. In physical applications
the space dimension $d+1$ is typically either 2 or 3. If one can solve the potential problem \eqref{lap}, then the vertical coordinate $z$ can be completely eliminated, so it is common to refer to $d$ as the dimension of the flow.

Taking the Fourier transform of the Laplace's equation with respect
to the horizontal coordinates results in a second-order ODE in $z$ whose solution enjoys the following integral
representation
\begin{equation}\label{eq2}
\phi({\bf x},z)=\frac{1}{\left(2\pi\right)^{d/2}}\int\hat{\phi}({\bf k})e^{qz}e^{i{\bf k}\cdot{\bf x}}d{\bf k},\ \ \ \hat{\phi}({\bf k})=\hat{\phi}^{*}(-{\bf k})
\end{equation}
consistent with the condition at infinity, that is, with vanishing
of ${\bf v}$ as $z\to-\infty$. Here ${\bf k}$ is the wave vector
and $q=|{\bf k}|$ (see Appendix \ref{sec:sec6} for the choice of Fourier transform
conventions, and other notations used throughout this manuscript).
Notice that the explicit time-dependence has been suppressed in \eqref{eq2}. 

Substitution of \eqref{eq2} into the expression for the kinetic energy followed by an explicit integration over $z$ yields \footnotesize
\[
K=\frac{1}{2}\int_{\mathbb{R}^{d}}d{\bf x}\int_{-\infty}^{\xi}\left(\left|\nabla\phi\right|^{2}+\phi_{z}^{2}\right)dz=\frac{1}{2\left(2\pi\right)^{d}}\int\frac{q_{0}q_{1}-{\bf k}\cdot{\bf k}_{1}}{q_{0}+q_{1}}\hat{\phi}_{0}\hat{\phi}_{1}e^{\left(q_{0}+q_{1}\right)\xi}e^{i\left({\bf k}+{\bf k}_{1}\right)\cdot{\bf x}}d{\bf k}_{01}d{\bf x}.
\] \normalsize
We pass to the weakly nonlinear limit by assuming that the surface
steepness is small, $\left|\nabla\xi\right|\ll1$, and develop
$e^{\left(q_{0}+q_{1}\right)\xi}$ in a Taylor series expansion around
the undisturbed surface level. The kinetic energy remains quadratic
in $\hat{\phi}$:
\begin{equation}\label{eq3}
K=\frac{1}{2}\int q_{0}\hat{\phi}_{0}\hat{\phi}_{0}^{*}d{\bf k}+\sum_{n=1}^{\infty}\int K_{0,1}^{(n+2)}\hat{\phi}_{0}\hat{\phi}_{1}\prod_{i=2}^{n+1}\hat{\xi}_{i}\delta^{(d)}\left(\sum_{i=0}^{n+1}{\bf k}_{i}\right)d{\bf k}_{012\dots n+1}
\end{equation}
where the kernels $K_{0,1}^{(n+2)}$ are computed straightforwardly
\[
K_{0,1}^{(n+2)}=\frac{q_{0}q_{1}-{\bf k}\cdot{\bf k}_{1}}{2\left(2\pi\right)^{nd/2}}\frac{\left(q_{0}+q_{1}\right)^{n-1}}{n!},\ \ \ n=1,2,\dots
\]
Note that $K_{0,1}^{(n+2)}=K_{1,0}^{(n+2)}$ for all $n$. 

Expressing $K$ in terms of surface variables alone is a laborious
task. One proceeds by finding a relation between the Fourier transform
of the velocity potential at the free surface, $\hat{\psi}\left({\bf k}\right)$,
and the variables $\hat{\phi}({\bf k}),\hat{\xi}({\bf k})$ using \eqref{eq2}:
\begin{equation}\label{eq4}
\hat{\psi}\left({\bf k}\right)=\hat{\phi}\left({\bf k}\right)+\sum_{n=1}^{\infty}\int\frac{q_{1}^{n}}{\left(2\pi\right)^{nd/2}n!}\hat{\phi}_{1}\prod_{i=2}^{n+1}\hat{\xi}_{i}\delta^{(d)}\left({\bf k}-\sum_{i=1}^{n+1}{\bf k}_{i}\right)d{\bf k}_{12\dots n+1}.
\end{equation}
Then inverts the relation \eqref{eq4} iteratively relative to $\hat{\phi}({\bf k})$,
inserts the result into \eqref{eq3}, collects terms of the same order in $\hat{\xi}({\bf k})$ and performs appropriate symmetrization.
These demanding calculations were carried out most comprehensively
by  \cite{K1} up to the fifth-order terms inclusive (see, also \cite{GAS, SS}). 

We shall take a slightly different path by postulating that 
\begin{equation}\label{eq5}
K=\frac{1}{2}\int q_{0}\hat{\psi}_{0}\hat{\psi}_{0}^{*}d{\bf k}+\sum_{n=1}^{\infty}\int E_{0,1,2,\dots,n+1}^{(n+2)}\hat{\psi}_{0}\hat{\psi}_{1}\prod_{i=2}^{n+1}\hat{\xi}_{i}\delta^{(d)}\left(\sum_{i=0}^{n+1}{\bf k}_{i}\right)d{\bf k}_{012\dots n+1}.
\end{equation}
Indeed, in the weakly non-linear regime the transformation $\left(\hat{\phi}_{{\bf k}},\hat{\xi}_{{\bf k}}\right)\to\left(\hat{\psi}_{{\bf k}},\hat{\xi}_{{\bf k}}\right)$
is a near-identity transformation as can be seen from  \eqref{eq4}, so we expect
the transformed kinetic energy to retain the same form as \eqref{eq3} except
that the kernels $K_{0,1}^{(n+2)}$ should be replaced by the kernels
$E_{0,1,2,\dots,n+1}^{(n+2)}$ depending, possibly, on all the wave
vectors participating in a $\left(n+2\right)$-wave interaction process.
In order to find these new kernels we compute the following variation
of the kinetic energy 

\begin{equation}\label{eq6}
\hat{\xi}_{t}({\bf k})=\frac{\delta K}{\delta\hat{\psi}_{0}^{*}}=q_{0}\hat{\psi}_{0}+2\sum_{n=1}^{\infty}\int E_{-0,1,2,\dots,n+1}^{(n+2)}\hat{\psi}_{1}\prod_{i=2}^{n+1}\hat{\xi}_{i}\delta^{(d)}\left({\bf k}-\sum_{i=1}^{n+1}{\bf k}_{i}\right)d{\bf k}_{12\dots n+1},
\end{equation}
and substitute \eqref{eq4} into the above expression. Comparing the result
to the Fourier transform of the kinematic boundary condition \eqref{eq1a} we
obtain (after proper symmetrization) the following expressions for the first three kermels:  

\begin{subequations}\label{eq7}
\begin{align}
2E_{0,1,2}^{(3)}= & -\frac{1}{\left(2\pi\right)^{d/2}}\left(q_{0}q_{1}+{\bf k}\cdot{\bf k}_{1}\right)\label{eq7a}\\
2E_{0,1,2,3}^{(4)}= & \frac{q_{1}}{2!\left(2\pi\right)^{d/2}}\left(E_{0,1,2+3}^{(3)}-E_{0,1+2,3}^{(3)}-E_{0,1+3,2}^{(3)}\right)+\left(0\longleftrightarrow1\right)\label{eq7b}\\
2E_{0,1,2,3,4}^{(5)}= & \frac{q_{1}^{2}}{3!\left(2\pi\right)^{d}}\left(E_{0,1,2+3+4}^{(3)}-E_{0,1+2+3,4}^{(3)}-E_{0,1+3+4,2}^{(3)}-E_{0,1+2+4,3}^{(3)}\right)\nonumber\\
 & -\frac{q_{1}}{3\left(2\pi\right)^{d/2}}\left(E_{0,1+2,3,4}^{(4)}+E_{0,1+3,2,4}^{(4)}+E_{0,1+4,2,3}^{(4)}\right)+\left(0\longleftrightarrow1\right)\label{eq7c}.
\end{align}
\end{subequations}
In general, before symmetrization we have the relation 
\[
2E_{0,1,2,\dots,n+1}^{(n+2)}+2\sum_{m=1}^{n-1}\frac{q_{1}^{m}E_{0,1+2+\dots+m+1,m+2,\dots,n+1}^{\left(n+2-m\right)}}{\left(2\pi\right)^{md/2}m!}=\frac{q_{1}^{n+1}+nq_{1}^{n-1}{\bf k}_{1}\cdot{\bf k}_{2}-q_{0}q_{1}^{n}}{\left(2\pi\right)^{nd/2}n!}
\] for $n\ge2$. 
Symmetrizing the right hand side of the above expression with respect to the wavenumbers $(2,3,\dots, n+1)$ yields $\frac{2E_{0,1,2+\dots+n+1}^{(3)}q_{1}^{n-1}}{\left(2\pi\right)^{(n-1)d/2}n!}$
thanks to the argument of the delta function appearing in \eqref{eq6}. This fact allows one to write a succinct formula for recursively determining all the kernels in terms of lower
order kernels 
\begin{equation}\label{eq8}
E_{0,1,2,\dots,n+1}^{(n+2)}=\frac{q_{1}^{n-1}E_{0,1,2+\dots+n+1}^{(3)}}{\left(2\pi\right)^{(n-1)d/2}n!}-\sum_{m=1}^{n-1}\frac{q_{1}^{m}E_{0,1+2+\dots+m+1,m+2,\dots,n+1}^{\left(n+2-m\right)}}{\left(2\pi\right)^{md/2}m!}
\end{equation}

The right hand side of  \eqref{eq8} should be symmetrized with respect to the pair of arguments $(2,3,\dots, n+1)$ and $(0,1)$ so that the total energy is symmetric. The advantage of the described method of finding the kernels is that it avoids the cumbersome operation of inverting  \eqref{eq4}. Following the outlined recipe a similar formula can be derived for waves in a fluid of arbitrary depth as shown in the Appendix \ref{sec:sec7}.

\subsection{\label{sec:sec2b} An application: Gravity waves in one horizontal dimension}

Two-dimensional ($d=1$) flows are special in many respects. It is well-know, for example, that using the power of complex variables theory the time-dependent fluid domain can be conformally mapped to a steady infinite strip or a half-plane \citep{DKSZ}. Obviously, this operation cannot be extended in general to higher dimensions. It appears that the dimension of the space plays a key role in our arguments too. Notice that the recursion
relation \eqref{eq8} is valid in all dimensions, however, the scalar product of two wave vectors
appearing in the kernel $E_{0,1,2}^{(3)}$ reduces to $k k_1$ only when $d=1$.
 
In this case it is also evident that $q_{0}q_{1}=\mbox{sgn}(k)\mbox{sgn}(k_{1})kk_{1}$
and therefore $E_{0,1,2}^{(3)}=0$ if and only if one of the wavenumbers
$k$ or $k_{1}$ is non-positive. We are not interested in wavenumbers
that are identically zero so they will be excluded from further consideration.
Let $\left\{ p_{j}\right\} _{j=1}^{n+1}$ be a set of arbitrary positive
integers (not necessarily distinct) such that $\sum_{j}p_{j}=p$.
Setting $k=-p$ and allowing each remaining $n+1$ wavenumbers to
assume any value from the set $\left\{ p_{1},p_{2},\dots,p_{n+1}\right\} $
without replacement (the set decreases when a wavenumber takes a certain
value), we find that 
\begin{equation}\label{eq9}
E_{0,1,2,\dots,n+1}^{(n+2)}=0
\end{equation} 
for all fixed $n\ge1$. Indeed, if $n=1$ then $E_{0,1,2}^{(3)}=0$ for an arbitrary choice
of $0<k_{1}\in\left\{ p_{1},p_{2}\right\} $ since $k=-(p_{1}+p_{2})$
is negative. To prove \eqref{eq9} for $n=2$ observe that, $E_{0,1,2+3}^{(3)}=E_{0,1+2,3}^{(3)}=E_{0,1+3,2}^{(3)}=0$
where now $k=-(p_{1}+p_{2}+p_{3})<0$ and each $k_{j}$ is assigned
a value from $\left\{ p_{1},p_{2},p_{3}\right\} $ without replacement.
We also have $E_{1,0+2,3}^{(3)}=E_{1,0+3,2}^{(3)}=0$ since $k+k_{j}<0$
for $j=2,3$ and hence $E_{0,1,2,3}^{(4)}=0$ according to \eqref{eq7b}. Similarly,
since the higher order kernels are given in terms of lower order kernels as in \eqref{eq8}
and $-k=\sum_{j}p_{j}$ is greater than all possible non-empty proper
subset sums of the set $\left\{ p_{j}\right\} _{j=1}^{n+1}$, we get
$E_{0,1,2,\dots,n+1}^{(n+2)}=0$ for an arbitrary but fixed $n\in\mathbb{N}$. The same reasoning can be applied to the set of arbitrary negative integers $\left\{ p_{j}\right\} _{j=1}^{n+1}$ to see that \eqref{eq9} still holds. 

For the kernels to vanish is it necessary to have $-k=\sum_{j}k_{j}$
with $k<0$ and $k_{j}>0$ for all $j$? One can find a lot of examples
where the latter condition is violated, but the kernel is still zero.
For instance, let $\{p_{1},p_{2},p_{3}\}=\{2,3,9\}$ and $k=-10$.
In this case $E_{0,1,2,3}^{(4)}=0$ but $\sum_{j}k_{j}\ne10$. However,
all such examples are flawed as far as the conservation of momentum
is concerned: every kernel in the expansion of the Hamiltonian is
multiplied by the delta function whose argument $k+\sum_{j}k_{j}$
expresses conservation of momentum. 

Shortly before Zakharov proposed the Hamiltonian formulation of water
waves, \citet{H1} developed the nonlinear energy transfer theory applicable
to all random wave fields including ocean waves  and adapted
Feynman's diagrammatic technique to compute transfer rates for various
scattering processes arising in geophysics \citep{H2}. Hasselmann
chose the convention that all interaction processes have a single
outgoing wave (cf. (1.29) and section 4 in \cite{H1}). This side condition pertains to our choice of wavenumbers (i.e. $k_j$ are all positive or negative and $-k=\sum_{j}k_{j}$) and \eqref{eq9} implies that waves propagating in one direction do not generate waves moving in the opposite direction.

It turns out that the same situation occurs in plasma physics: the absence of nonlinear interaction between co-propagating Alfv\'{e}n waves in incompressible MHD has been recently reported in \cite{HN} without appealing to the Hamiltonian formalism. 

The remarkable utility of the recursion relation \eqref{eq8} is that it offers
a simple way of arriving at \eqref{eq9} without even knowing the explicit expressions
for the individual kernels. The three-wave interaction kernel can be used as a building block for the construction of the higher order kernels via the recursion. For the sake of completeness below we record the explicit forms of a few of the kernels computed using \eqref{eq8}:  

\begin{align*} 
E_{0,1,2}^{(3)}= & -\frac{1}{2\left(2\pi\right)^{d/2}}\left(q_{0}q_{1}+{\bf k}\cdot{\bf k}_{1}\right)\\
E_{0,1,2,3}^{(4)}= & -\frac{1}{8\left(2\pi\right)^{d}}\left(2|{\bf k}|^{2}q_{1}+2|{\bf k}_{1}|^{2}q_{0}-q_{0}q_{1}\left(q_{0+2}+q_{0+3}+q_{1+2}+q_{1+3}\right)\right)\\
E_{0,1,2,3,4}^{(5)}= & -\frac{1}{12\left(2\pi\right)^{3d/2}}\left(-|{\bf k}_{1}|^{2}q_{0}\left(q_{0+2}+q_{0+3}+q_{0+4}\right)-|{\bf k}|^{2}q_{1}\left(q_{1+2}+q_{1+3}+q_{1+4}\right)\right. \\ & \left.+2|{\bf k}|^{2}|{\bf k}_{1}|^{2}+\frac{q_{0}q_{1}}{2}\left(|{\bf k}|^{2}-|{\bf k}+{\bf k}_{2}|^{2}-|{\bf k}+{\bf k}_{3}|^{2}-|{\bf k}+{\bf k}_{4}|^{2}\right)\right. \\ & \left.+\frac{q_{0}q_{1}}{2}\left(|{\bf k}_{1}|^{2}-|{\bf k}_{1}+{\bf k}_{2}|^{2}-|{\bf k}_{1}+{\bf k}_{3}|^{2}-|{\bf k}_{1}+{\bf k}_{4}|^{2}\right)\right. \\ &
	\left.+q_{0}q_{1}\left(q_{0+2}\left(q_{1+3}+q_{1+4}\right)+q_{0+3}\left(q_{1+2}+q_{1+4}\right)+q_{0+4}\left(q_{1+2}+q_{1+3}\right)\right)\right)
\end{align*} 
These expressions are in full agreement with those previously obtained by \citet{K1} and other authors, e.g., \cite{GAS, SS}. Given their increasing complexity, it is highly unlikely that \eqref{eq9} could have been obtained directly. 

The Hamiltonian truncated to a finite order in slope induces approximate systems that are capable of retaining, at least to a certain extent, all the important characteristics of the evolution. The equation of motion corresponding to the Hamiltonian containing only quadratic and quartic terms in the normal variables (see, \eqref{rdH})  is called the Zakharov equation. Its properties have been studied in great detail, however, the integrability is still an open problem.
In the Introduction we mentioned the one dimensional nonlinear Sch\"{o}dinger equation (NLS). It is the first and the only known completely integrable model approximating deep-water surface waves. Yet the NLS equation is not specific just to the surface waves on deep water - it commonly appears in other nonlinear dispersive energy-preserving systems and  can be derived from, e. g., the nonlinear Klein-Gordon and the KdV equations \citep{A}. By virtue of \eqref{eq8} calculations that used to be infeasible become more tractable and the prospect of exploring new integrable systems and/or new integrals of motion seems promising.
Furthermore, the possible role of \eqref{eq8} in numerical implementation of the evolution equations can hardly be exaggerated since the kernels $E_{0,1,2,\dots,n+1}^{(n+2)}$ appear not only in the expansion of the Hamiltonian, but also in the evolution of $\hat{\xi}({\bf k})$ and $\hat{\psi}({\bf k})$ (see, \eqref{eq6}).

\section{\label{sec:sec3} Relationship with the existing literature}

The Hamiltonian description provides a unified framework for studying
various types of waves without appealing to the specifics of the medium
thus treating surface and spin waves, waves in plasma and nonlinear
optics on an equal footing. This universality can be attributed, in
part, to the fact that linear waves satisfying a given dispersion
relation $\omega\left({\bf k}\right)$ exhibit the same behavior regardless
whether they propagate, for instance, in plasma, fluid or on a string.
For a given medium the dispersion relation is determined from the
linearization of the equations of motion around the rest state and
in particular for deep water gravity waves takes the form $\omega({\bf k})=\sqrt{g|{\bf k}|}$
(see \eqref{eq1}). Another important advantage of the Hamiltonian formalism
is that it allows us to employ a wide class of field transformations
preserving the canonical structure of the equations of motion. The
most basic example of such a transformation was given by \citet{Z}
who defined the wave amplitude $a({\bf k})$ and its complex conjugate
related to the physical variables $\hat{\psi}_{{\bf k}}$ and $\hat{\xi}_{{\bf k}}$
by 
\begin{equation}\label{eq10}
\hat{\xi}\left({\bf k}\right)=\sqrt{\frac{q\left({\bf k}\right)}{2\omega\left({\bf k}\right)}}\left(a({\bf k})+a^{*}(-{\bf k})\right)\mbox{ and }\hat{\psi}\left({\bf k}\right)=-i\sqrt{\frac{\omega\left({\bf k}\right)}{2q\left({\bf k}\right)}}\left(a({\bf k})-a^{*}(-{\bf k})\right).
\end{equation}
The variables $ia^{*}({\bf k})$ and $a({\bf k})$ are canonically
conjugate, and in accordance with the wave-particle duality could
be considered as creating and respectively annihilating a wave of
momentum ${\bf k}$ and energy $\omega({\bf k})$. Using the linear
transformation \eqref{eq10} it is possible to write the Hamiltonian of the system
(see \eqref{eq5}) as a series expansion in integer powers of $a$ and $a^{*}$,
grouping the terms as 
\begin{equation}\label{ham}
H(a,a^{*})=H_{2}+H_{int}=\int\omega\left({\bf k}\right)a^{*}({\bf k})a\left({\bf k}\right)d{\bf k}+H_{int}
\end{equation}
where $H_{2}$
is the Hamiltonian of the linearized theory and the interaction Hamiltonian
$H_{int}$ is a sum over all terms of the form \footnotesize
\[
\int V^{(m,n)}\left({\bf k}_{1},{\bf k}_{2},\dots,{\bf k}_{n+m}\right)a_{1}^{*}a_{2}^{*}\dots a_{m}^{*}a_{m+1}\dots a_{n+m}\delta^{(d)}\left(\sum_{i=1}^{m}{\bf k}_{i}-\sum_{i=m+1}^{n+m}{\bf k}_{i}\right)d{\bf k}_{12\dots n+m},
\] \normalsize
with $n+m\geq3$, representing a wave-wave interaction process where $m$ waves are
created and $n$ waves are annihilated. Insisting that $H$ remains
real and symmetric under relabeling of dummy integration variables,
we impose certain constraints on the kernels. For example, in the problem
under consideration the kernel $V^{(2,2)}\left({\bf k}_{1},{\bf k}_{2},{\bf k}_{3},{\bf k}_{4}\right)$
should satisfy $V_{1,2,3,4}^{(2,2)}=V_{2,1,3,4}^{(2,2)}=V_{1,2,4,3}^{(2,2)}=V_{3,4,1,2}^{(2,2)}$. 

The ``free'' Hamiltonian $H_{2}$ can be associated with the energy
of a collection of non-interacting harmonic oscillators generalized
to the case of a continuum. The evolution equation for $a\left({\bf k}\right)$
reads

\begin{equation}\label{eq11}
i\frac{\partial a\left({\bf k}\right)}{\partial t}-\omega\left({\bf k}\right)a\left({\bf k}\right)=\frac{\delta H_{int}}{\delta a^{*}\left({\bf k}\right)}
\end{equation}

The right-hand side of \eqref{eq11} acts as a driving force and can induce a
resonant interaction between waves if the frequency of a ``free''
wave coincides with that of the source term manifesting itself through
products of amplitudes. In other words, a significant energy transfer
among $n$ waves of different wavelengths can occur if the dispersion relation $\omega({\bf k})$
admits nontrivial pairs $\left({\bf k}_{i},\omega\left({\bf k}_{i}\right)\right)$
simultaneously obeying the resonance conditions 

\begin{subequations}\label{eq12}
\begin{align}
\omega({\bf k}_{1})\pm\omega({\bf k}_{2})\pm\dots\pm\omega({\bf k}_{n}) & =  0,\label{eq12a}\\
{\bf k}_{1}\pm{\bf k}_{2}\pm\dots\pm{\bf k}_{n} & =  0\label{eq12b}
\end{align}
\end{subequations}
for some choice of signs. These condition correspond
to conservation of energy and momentum in the wave-particle analogy. 

As a consequence of the upward convexity
of the function $\omega({\bf k})=\sqrt{g|{\bf k}|}$, i.e. $d^{2}\omega/d|{\bf k}|^{2}<0$,
resonant triads are not possible so we are motivated to investigate
the four-wave interactions \citep{P}. Three different kinds of quartic interactions
should be analyzed: i) the process of the annihilation/creation of
four waves, labeled $4\leftrightarrow0$, corresponding to the case
when all signs in \eqref{eq12} are $+1$; ii) the process in which three waves
combine into one or one wave decays into three, $3\leftrightarrow1$,
representing the case when all but one term in  \eqref{eq12} has the minus sign;
and iii) the $2\leftrightarrow2$ process in which two waves decay/combine
into two waves (two terms in \eqref{eq12} have negative signs). It turns out
that only the latter process is resonant and there are no resonant
four-wave interactions of the type $4\leftrightarrow0$ and $3\leftrightarrow1$.
As a matter of fact, in isotropic media the resonance conditions 

\begin{subequations}\label{eq13}
\begin{align}
\omega({\bf k}_{1})+\omega({\bf k}_{2}) & =\omega({\bf k}_{3})+\omega({\bf k}_{4})\\
{\bf k}_{1}+{\bf k}_{2} & ={\bf k}_{3}+{\bf k}_{4}
\end{align}
\end{subequations}
are satisfied irrespective of the shape of the dispersion curve whenever
$d\geq2$. Solutions to \eqref{eq13}  include degenerate cases when all the wavenumbers
are equal or when ${\bf k}_{1}={\bf k}_{3}$ and ${\bf k}_{2}={\bf k}_{4}$
up to permutations of the indices $3$ and $4$. These generic interactions
are not very interesting because they result only in a frequency shift,
but no exchange of energy \citep{D}. 

When $d=1$, the resonant conditions \eqref{eq13} are verified by a family of
wavenumbers and frequencies 

\begin{subequations}\label{eq14}
\begin{align}
 & \left(k_{1},k_{2},k_{3},k\right)=\alpha\times\left((\zeta+1)^{2},\zeta^{2}(\zeta+1)^{2},-\zeta^{2},(\zeta^{2}+\zeta+1)^{2}\right)\\
 & \left(\omega_{1},\omega_{2},\omega_{3},\omega\right)=\sqrt{g|\alpha|}\times\left(\zeta+1,\zeta(\zeta+1),\zeta,\zeta^{2}+\zeta+1\right)
\end{align}
\end{subequations}
where $0<\zeta\leq1$ and $\alpha\ne0$ are chosen so that each $k_{i}$
is an integer \citep{DZ1}. 

The Hamiltonian  \eqref{ham} expressed in terms of $a$ and $a^{*}$ is, in a sense,
not optimal because it contains non-resonant cubic and quartic terms
making the right-hand side of \eqref{eq11} unnecessarily complicated. A considerable
simplification of the nonlinear evolution equation can be achieved
by eliminating these unimportant terms by one way or another. A widely
exploited method for deriving the simplified Hamiltonian $\tilde{H}$
is to perform a suitable near identity transformation from $a$ and
$ia^{*}$ to a new set of canonically conjugate variables $b$ and
$ib^{*}$. Unlike \eqref{eq10}, the transformation $a=a\left(b,b^{*}\right)$
is non-linear and postulated as a series expansion in integer powers
of $b$ and $b^{*}$ with the expansion coefficients satisfying certain
relations ensuring that the transformation is canonical up to a desired
order (see, e.g., \cite{K1} for details). If the transformation is canonical up to
terms of order four inclusive, then the reduced Hamiltonian reads 
\begin{align} \label{rdH}
\tilde{H}(b^{*},b) & = \int\omega_{0}b_{0}^{*}b_{0}d{\bf k}+\int T_{1,2,3,4}^{(2,2)}b_{1}^{*}b_{2}^{*}b_{3}b_{4}\delta_{1+2-3-4}d{\bf k}_{1234}\\ &+\int T_{1,2,3,4,5}^{(2,3)}\left(b_{1}^{*}b_{2}^{*}b_{3}b_{4}b_{5}+\mbox{c.c.}\right)\delta_{1+2-3-4-5}d{\bf k}_{12345}+R_{6} \nonumber
\end{align}  
where $R_{6}$ is a remainder containing degree six and higher terms
both resonant and non-resonant and  c.c. stands for complex conjugate. The non-resonant terms can be removed order by order systematically, should the need arise. 

We mentioned in the Introduction that vanishing of the four-wave interaction kernel
$T_{1,2,3,4}^{(2,2)}$ on the resonant
hypersurface \eqref{eq13} takes place when $d=1$. This implies that only trivial scattering can occur in a four-wave nonlinear interaction processes on deep water. Moreover, when $d=1$ Lvov found that the five-wave interaction kernel $T_{1,2,3,4,5}^{(2,3)}$ is zero on the resonant hypersurface 

\begin{subequations}\label{eq15}
\begin{align}
\omega({\bf k}_{1})+\omega({\bf k}_{2}) & =\omega({\bf k}_{3})+\omega({\bf k}_{4})+\omega({\bf k}_{5})\\
{\bf k}_{1}+{\bf k}_{2} & ={\bf k}_{3}+{\bf k}_{4}+{\bf k}_{5}
\end{align}
\end{subequations}
whenever i) $k_{1}k_{2}<0$ and $k_{3},k_{4},k_{5}$ are of the same
sign or ii) when $k_{1}k_{2}>0$ and one of the wavenumbers  $k_{3},k_{4},k_{5}$
has the same sign as $k_{1}$ and $k_{2}$ \citep{L}. Despite the very lengthy
intermediate calculations, the final expressions for $T_{1,2,3,4,5}^{(2,3)}$
produced by all other sign combinations (up to relabeling of the wavenumbers)
are astonishingly compact as shown in \cite{DLZ, L}. 

A parametrization of the solution set of \eqref{eq15} corresponding to the vanishing
$T_{1,2,3,4,5}^{(2,3)}$, to the best of our knowledge, has not been previously reported so we present
it in what follows. On the hypersurface \eqref{eq15} the five-wave interaction amplitude
is zero as long as 

\footnotesize
\begin{subequations}\label{eq16}
\begin{align}
 & \left(k_{1},k_{2},k_{3},k_{4},k_{5}\right)=\alpha\times\left(-\left(\zeta_{+}^{2}-\zeta^{2}\right)^{2},\left(\zeta_{+}^{2}+\zeta^{2}\right)^{2},(2\zeta_{1}\zeta_{+})^{2},(2\zeta_{2}\zeta_{+})^{2},(2\zeta_{3}\zeta_{+})^{2}\right)\\
 & \left(\omega_{1},\omega_{2},\omega_{3},\omega_{4},\omega_{5}\right)=\sqrt{g|\alpha|}\times\left(\zeta_{+}^{2}-\zeta^{2},\zeta_{+}^{2}+\zeta^{2},2\zeta_{1}\zeta_{+},2\zeta_{2}\zeta_{+},2\zeta_{3}\zeta_{+}\right)
\end{align}
\end{subequations} \normalsize
or 

\footnotesize
\begin{subequations}\label{eq17} 
\begin{align}
 & \left(k_{1},k_{2},k_{3},k_{4},k_{5}\right)=\alpha\times\left((2\zeta_{1}\zeta_{-})^{2},(2\zeta_{2}\zeta_{-})^{2},-(2\zeta_{3}\zeta_{-})^{2},-(\zeta_{-}^{2}-\zeta^{2})^{2},(\zeta_{-}^{2}+\zeta^{2})^{2}\right)\\
 & \left(\omega_{1},\omega_{2},\omega_{3},\omega_{4},\omega_{5}\right)=\sqrt{g|\alpha|}\times\left(2\zeta_{1}\zeta_{-},2\zeta_{2}\zeta_{-},2\zeta_{3}\zeta_{-},\zeta_{-}^{2}-\zeta^{2},\zeta_{-}^{2}+\zeta^{2}\right)
\end{align}
\end{subequations} \normalsize
where $\zeta^2=\sum_{i=1}^3\zeta_{i}^{2}$, $\zeta_{\pm}=\zeta_{1}+\zeta_{2}\pm\zeta_{3},$ the scaling factor $\alpha$ and the parameters $\zeta_{1},\zeta_{2},\zeta_{3}$
are real, non-zero numbers such that each $k_{i}$ is an integer and
each $\omega_{i}$ is positive. For a specific choice of these parameters one can recover resonant quintets considered by \cite{BS} in their study of existence, stability, and long-time behavior of time-periodic standing waves on deep water. 

How is the vanishing of $T_{1,2,3,4}^{(2,2)}$ and $T_{1,2,3,4,5}^{(2,3)}$
on the hypersurfaces \eqref{eq13} and \eqref{eq15} related to the vanishing of $E_{0,1,2,3}^{(4)}$
and $E_{0,1,2,3,4}^{(5)}$ given that any canonical transformation
performed on the Hamiltonian $H$ does not influence the way waves
propagate and interact with each other? In Sec. \ref{sec:sec2b} we proved that $E_{0,1,2,\dots,n+1}^{(n+2)}=0$
when $n+1$ wavenumbers $k_{1},\dots,k_{n+1}$ are sampled without
repetition from the set $\{p_{1},p_{2},\dots,p_{n+1}\}$ of positive
integers satisfying $\sum_{j}p_{j}=p$ and $k$ is set to be $-p$.
Clearly, the statement holds true if we let $p_{j}=l_{j}^{2}$. Now,
if we restrict the sum $\sum_{j}l_{j}^{2}$ to be a complete square,
say $l^{2}$, then finding parameterizations such as \eqref{eq14}, \eqref{eq16} and \eqref{eq17} becomes
an exercise in the theory of quadratic forms. Let us, to be definite,
consider the equation
\begin{equation}\label{eq18}
l_{1}^{2}+l_{2}^{2}+l_{3}^{2}+l_{4}^{2}=l^{2}.
\end{equation}
The dispersion law vetoes interactions of the type $5\leftrightarrow0$
and $4\leftrightarrow1$, and we are left with the $2\leftrightarrow3$
interaction. Up to combinatorial equivalence there are two ways to
arrange the terms in \eqref{eq18} so that the resulting equation corresponds
to the conservation of momentum in the $2\leftrightarrow3$ process.
Crossing, say $l_{4}^{2}$, over to the right-hand side of \eqref{eq18} and constructing
a solution consistent with $l_{1}+l_{2}+l_{3}=l+l_{4}$, we obtain
the parameterization \eqref{eq16} identifying $k_{1}=l^{2}$, $k_{2}=-l_{4}^{2}$,
$k_{3}=l_{3}^{2}$, $k_{4}=l_{2}^{2}$, $k_{5}=l_{1}^{2}$ up to scaling with a constant. Similarly,
crossing $l_{4}^{2}$ and $l_{3}^{2}$ over and constructing a solution
consistent with $l_{1}+l_{2}=l+l_{3}+l_{4}$, we get \eqref{eq17} identifying
$k_{1}=l_{1}^{2}$, $k_{2}=l_{2}^{2}$, $k_{3}=-l_{3}^{2}$, $k_{4}=-l_{4}^{2}$,
$k_{5}=l^{2}$. Thus by examining all Pythagorean 5-tuples $(l_{1},l_{2},l_{3},l_{4},l)$
compatible with the conservation of energy we found two parameterizations
of the resonant hypersurface \eqref{eq15} on which the five-wave interaction kernel $T_{1,2,3,4,5}^{(2,3)}$
is zero. The resonant hypersurface \eqref{eq15} admits other solutions,
for example, those in which all the wave numbers $k_{1}$ through
$k_{5}$ are positive. In this case, we can parameterize the hypersurface
by considering the relation $l_{1}+l_{2}+l_{3}=l+l_{4}$ and $l_{1}^{2}+l_{2}^{2}+l_{3}^{2}-l_{4}^{2}=l^{2}$ instead of \eqref{eq18}. A rather lengthy calculation
results in a remarkably compact expression for the interaction kernel
which yields non-zero values on the hypersurface. Observe that in this case $E_{-l^{2},l_{1}^{2},l_{2}^{2},l_{3}^{2},-l_{4}^{2}}^{(5)}\ne0$ as well. 

In the normal variables representation, the number of terms contributing
to the calculation of the higher order (six and above) interaction
amplitudes presents a huge analytical and numerical obstacle even
after resorting to symbolic manipulators such as Maple and Mathematica.
Without deriving the explosively large expressions for the interaction
amplitudes, is it possible to obtain some non-trivial configurations
of the wave vectors for which these interaction kernels vanish? Let
us explore this possibility for the six-wave interaction processes.
The dispersion relation allows two types of resonant interactions,
$2\leftrightarrow4$ and $3\leftrightarrow3$, and the associated
resonance conditions are 

\begin{subequations}\label{res1} 
\begin{align}
\omega({\bf k}_{1})+\omega({\bf k}_{2}) & =\omega({\bf k}_{3})+\omega({\bf k}_{4})+\omega({\bf k}_{5})+\omega({\bf k}_{6}) \label{res1a} \\ {\bf k}_{1}+{\bf k}_{2} & ={\bf k}_{3}+{\bf k}_{4}+{\bf k}_{5}+{\bf k}_{6} \end{align} \end{subequations}
and 
\begin{subequations}\label{res2}
\begin{align}
 \omega({\bf k}_{1})+\omega({\bf k}_{2})+\omega({\bf k}_{3}) & =\omega({\bf k}_{4})+\omega({\bf k}_{5})+\omega({\bf k}_{6}) \label{res2a} \\ 
{\bf k}_{1}+{\bf k}_{2}+{\bf k}_{3} & ={\bf k}_{4}+{\bf k}_{5}+{\bf k}_{6}
\end{align} \end{subequations}
Without performing explicit computations we can deduce that the six-wave
interaction kernel $T_{1,2,3,4,5,6}^{(2,4)}$ is zero on the
hypersurface \eqref{res1} in $d=1$ whenever there are wavenumbers such that i) $k_{1}k_{2}<0$ and $k_{3},k_{4},k_{5},k_{6}$
are of the same sign or ii) $k_{1}k_{2}>0$ and only one
of the wavenumbers  $k_{3},k_{4},k_{5},k_{6}$ has the same sign as
$k_{1}$, and simultaniously respecting the conservation of energy \eqref{res1a}. Furthermore, the interaction kernel $T_{1,2,3,4,5,6}^{(3,3)}$
is zero on the hypersurface \eqref{res2} when $k_{1}k_{2}<0$, $k_{1}k_{3}<0$
and $k_{4},k_{5},k_{6}$ are of the same sign as $k_{1}$, again up to relabelling of the wavenumbers and obeying \eqref{res2a}. This is
just an obfuscated way of presenting a simple fact that in the $(\hat{\psi},\hat{\xi})$-representation
$E_{-l^{2},l_{1}^{2},l_{2}^{2},l_{3}^{2},l_{4}^{2},l_{5}^{2}}^{(6)}=0$
for any Pythagorean 6-tuple $(l_{1},l_{2},l_{3},l_{4},l_{5},l)$ without regard to the constraint on the frequencies. It may happen that for some dispersion relation and order of approximation the interaction kernel is zero in the $(b,b^*)$-representation irrespective of the frequencies (cf. the five-wave amplitude for the KdV equation in \citep{ZOCO}), however, such an occurrence is exceptional and cannot even be guaranteed to carry over to the next order in the expansion.

Introducing the normal variables reduces the number of equations of motion from two coupled equations for $\hat\xi$ and $\hat\psi$ to one complex-valued $\eqref{eq11}$, but brings ambiguity in the relative orientations of the wave vectors. Nevertheless, the whole procedure of singling out those orientations for which the interaction amplitudes are zero can be made straightforward. If the dispersion relation tells us that a certain $m\leftrightarrow n$ process is resonant, then to find parameterization(s) of the resonant hypersurface corresponding to the vanishing amplitude we should first regroup the wavenumbers in such a way as to have one outgoing and $n+m-1$ incoming waves with all positive or all negative wavenumbers obeying conservation of momentum (as in \eqref{eq18} for five waves), and look for possible non-trivial solutions compatible with the condition on the frequencies. 

\section{\label{sec:sec4} Discussion}

The main technical challenge posed in the Hamiltonian formulation of surface waves concerns expressing the kinetic energy as a functional of the surface elevation and the velocity potential at the free surface. Nearly all known explicit results are given as a power series in the surface elevation. The first three and four terms of the expansion were obtained by \cite{Z} and by \cite{K1} respectively, the generalization to an arbitrary order was addressed by \citet{GAS}. These derivations require a series reversion at an intermediate stage of the calculation. A practical advantage of the approach presented in this article is that it avoids the cumbersome operation of inversion and results in a recursion relation that is convenient for the analysis of higher order wave-wave interaction amplitudes. In fairness, it should be mentioned that there is another method which does not involve a series reversion, but it relies on a reformulation of the theory in terms of a non-local Dirichlet-to-Neumann (DtN) operator relating the values of the velocity potential at the free surface to the surface values of its normal derivative \citep{CS}. Yet the recursion derived from the DtN formulation is not best suited for the interaction picture and classical scattering matrix theory. 

Switching from the physical variables, the free surface elevation
and velocity potential, to the complex wave amplitudes allows one
to pursue the particle analogy in which interactions between waves
are regarded as particle collisions. Defining the
correlation function $\left\langle a({\bf k}),a^{*}({\bf k}')\right\rangle $
facilitates the derivation of the so-called kinetic equation for surface
waves which has a form of the classical Boltzmann equation
describing the evolution of the density of interacting particles 
of momentum ${\bf k}$ and energy $\omega$
satisfying the energy-momentum (dispersion) relation \citep{K2}. However, the
transition $(\hat{\psi},\hat{\xi})\to(a,ia^{*})$ inevitably generates
more terms in the expansion of the Hamiltonian naturally leading to
more time-consuming computations. The terms can be grouped as  $H_{3}(\hat{\xi},\hat{\psi})=H_{3}^{3\leftrightarrow0}(a^{*},a)+H_{3}^{2\leftrightarrow1}(a^{*},a)$,
$H_{4}(\hat{\xi},\hat{\psi})=H_{4}^{4\leftrightarrow0}(a^{*},a)+H_{4}^{3\leftrightarrow1}(a^{*},a)+H_{4}^{2\leftrightarrow2}(a^{*},a)$
and so on, prompting the need for a further canonical transformation
$(a,ia^{*})\to(b,ib^{*})$ which excludes the non-resonant interactions.
On the other hand, the underlying physics should not be affected by
a canonical transformation and if waves do not interact, that is, if
a certain interaction kernel of the original Hamiltonian vanishes
for some configuration of the wave vectors, then the same should also
hold for the corresponding kernel of a reduced Hamiltonian $\tilde H(b,b^*)$ provided
that the conservation laws are respected. In Sec. \ref{sec:sec3} we have verified
this claim for the four- and five-wave interaction amplitudes by crossing
some of the wavenumbers satisfying the resonance condition over to
the other side of the equation. 
In particle physics
this principle is known as crossing symmetry \citep{PS}. Roughly, it states that
the amplitude for any process involving a particle with the incoming
momentum ${\bf k}$ is equal to the amplitude for a similar process
but with an antiparticle of outgoing momentum $-{\bf k}$. Thus we can assume that there is only one outgoing wave in any interaction process. In Hasselmann's terminology there is a "passive" wave component, having no direct influence on the interaction, which receives energy from the "active" wave components \citep{H1}. One should
bear in mind that the conservation of energy can prohibit a process
which is otherwise permissible (for example, a lighter particle cannot
decay into a heavier particle). Analogously, the shape of the dispersion curve tells us which wave-wave interaction processes are allowed and which should be ruled out. 

\acknowledgements
The author is supported by the KAUST Fellowship. 

\appendix
\section{\label{sec:sec6} Conventions}
Throughout this article we treat the fluid density as constant in both space and time by setting $\rho=1$ and neglect effects due to surface tension. The vertical coordinate $z$ points upwards and the undisturbed fluid surface coincides with the hyperplane $z=0$. The Fourier transform of a given function $f({\bf x})$, the inverse transform and the Dirac delta function are defined as follows 
\footnotesize
$$\hat f({\bf k})=\frac{1}{\left(2\pi\right)^{d/2}}\int{f}({\bf x})e^{-i{\bf k}\cdot{\bf x}}d{\bf x}, \ \ f({\bf x})=\frac{1}{\left(2\pi\right)^{d/2}}\int\hat{f}({\bf k})e^{i{\bf k}\cdot{\bf x}}d{\bf k}, \ \ \delta^{(d)}({\bf k})=\frac{1}{\left(2\pi\right)^{d}}\int e^{i{\bf k}\cdot{\bf x}}d{\bf x}.$$ \normalsize
We use the shorthand notation in which the arguments of functions are replaced by subscripts, for example, $\omega_j=\omega({\bf k}_j)$, $\hat\xi_j=\hat\xi({\bf k}_j,t)$, $\delta_{0-1-2}=\delta({{\bf k}-{\bf k}_{1}-{\bf k}_{2}})$, etc. where subscript 0 corresponds to $\bf k$. Multiple integrals are represented by a single integral sign with a differential being the appropriate volume element, e.g., $d{\bf k}_{012}$ stands for $d{\bf k}d{\bf k}_{1}d{\bf k}_{2}$.  
\section{\label{sec:sec7} Derivation of the recursion relation for  an arbitrary depth} 
Assume that the fluid depth is bounded from below, namely, let a solid bottom be located at constant depth $z=-h$. Then the solution to the Laplace equation $\eqref{lap}$ satisfying the impermeability condition $\phi_z=0$ at the bottom boundary reads
\[
\phi({\bf x},z)=\frac{1}{(2\pi)^{d/2}}\int\hat{\phi}({\bf k})\frac{\cosh\left(|{\bf k}|(z+h)\right)}{\cosh\left(|{\bf k}|h\right)}e^{i{\bf k}\cdot{\bf x}}d{\bf k}, \ \ \ \hat{\phi}({\bf k})=\hat{\phi}^{*}(-{\bf k}).
\] 
Evaluating $\phi({\bf x},z)$ at the free surface $z=\xi({\bf x},t)$ in the weakly non-linear approximation we find 

\begin{equation}\label{eq19}
\hat{\psi}\left({\bf k}\right)=\hat{\phi}\left({\bf k}\right)+\sum_{n=1}^{\infty}\int\frac{|{\bf k}_{1}|^{n}\mu_n(|{\bf k}_{1}|)}{\left(2\pi\right)^{nd/2}n! \mu_0(|{\bf k}_{1}|)}\hat{\phi}_{1}\prod_{i=2}^{n+1}\hat{\xi}_{i}\delta^{(d)}\left({\bf k}-\sum_{i=1}^{n+1}{\bf k}_{i}\right)d{\bf k}_{12\dots n+1}.
\end{equation}
where $\mu_n(|{\bf k}|)=e^{|{\bf k}|h}+(-1)^n e^{-|{\bf k}| h}$. Notice that for a fixed $n\in\mathbb{N}$ the ratio $\mu_n(|{\bf k}|)/\mu_0(|{\bf k}|)$ is either equal to 1 or tends to 1 as $h\to \infty$ and $\eqref{eq19}$ reduces to $\eqref{eq4}$ in this limit. Substituting  $\eqref{eq19}$ into $\eqref{eq5}$ and comparing the result to the Fourier transform of the kinematic boundary condition  $\eqref{eq1a} $ we arrive at the desired formula

\begin{subequations}
\begin{align*} 
E_{0,1,2}^{(3)}= &  -\frac{1}{2\left(2\pi\right)^{d/2}}\left(q_{0}q_{1}+{\bf k}\cdot{\bf k}_{1}\right), \\
 E_{0,1,2,\dots,n+1}^{(n+2)}= &   \frac{|{\bf k}_{1}|^{n+1}\mu_{n+1}(|{\bf k}_{1}|)+n{\bf k}_{1}\cdot{\bf k}_{2}|{\bf k}_{1}|^{n-1}\mu_{n-1}(|{\bf k}_{1}|)-q_{0}|{\bf k}_{1}|^n\mu_n(|{\bf k}_{1}|)}{2\left(2\pi\right)^{nd/2}n!\mu_0(|{\bf k}_{1}|)}\\ &-\sum_{m=1}^{n-1}\frac{|{\bf k}_{1}|^{m}\mu_m(|{\bf k}_{1}|)E_{0,1+2+\dots+m+1,m+2,\dots,n+1}^{\left(n+2-m\right)}}{\left(2\pi\right)^{md/2}m!\mu_0(|{\bf k}_{1}|)}, \  n=2, 3,\dots 
\end{align*}
\end{subequations} 
where $q({\bf k})=|{\bf k}|\tanh({|{\bf k}|h})$ and depending whether $n\geq 2$ is odd or even a further simplification is possible since $|{\bf k}_{1}|\mu_n(|{\bf k}_1|)/\mu_0(|{\bf k}_1|)$ can be replaced by $q_1$ or $|{\bf k}_{1}|$ respectively. In the limit of an infinitely deep fluid the above expression reduces to $\eqref{eq8}$.

\end{document}